\documentclass[showkeys,amsmath,amssymb,twocolumn,prb]{revtex4}
\usepackage{amsmath,amsthm}
\usepackage{graphicx}
\usepackage{dcolumn}
\usepackage{bm}
\usepackage{color,epsfig,multirow}
\newcommand{\gan}[1] {{\color{red} #1 }}
\newcommand{\xing}[1] {{\color{green} #1 }}

\begin{document}


\title{Effects of image charges, interfacial charge discreteness, and surface roughness
on the zeta potential of spherical electric double layers}

\author{Zecheng Gan$^{1,3}$}\email{ganzecheng1988@sjtu.edu.cn}
\author{Xiangjun Xing$^{2,3}$}\email{xxing@sjtu.edu.cn}
\author{Zhenli Xu$^{1,3}$}\email{xuzl@sjtu.edu.cn}
\affiliation{1. Department of Mathematics, Shanghai Jiao Tong University, Shanghai 200240, China}
\affiliation{2. Department of Physics, Shanghai Jiao Tong University, Shanghai 200240, China}
\affiliation{3. Institute of Natural Sciences, Shanghai Jiao Tong University, Shanghai 200240, China}

\date{\today}

\begin{abstract}
We investigate the effects of image charges, interfacial charge discreteness, and surface roughness on spherical electric double layer structures in electrolyte solutions with divalent counter-ions in the setting of the primitive model. By using Monte Carlo simulations and the image charge method, the zeta potential profile and the integrated charge distribution function are computed for varying surface charge strengths and salt concentrations.  Systematic comparisons were carried out between three distinct models for interfacial charges: 1) SURF1 with uniform surface charges, 2) SURF2 with discrete point charges on the interface, and 3) SURF3 with discrete interfacial charges and finite excluded volume.  By comparing the integrated charge distribution function and the zeta potential profile, we argue that the potential at the distance of one ion diameter from the macroion surface is a suitable location to define the zeta potential.  In SURF2 model, we find that image charge effects strongly enhance charge inversion for monovalent interfacial charges, and strongly suppress charge inversion for multivalent interfacial charges.  For SURF3, the image charge effect becomes much smaller. Finally, with image charges in action, we find that excluded volumes (in SURF3) suppress charge inversion for monovalent interfacial charges and enhance charge inversion for multivalent interfacial charges.  Overall, our results demonstrate that all these aspects, i.e., image charges, interfacial charge discreteness, their excluding volumes have significant impacts on zeta potentials of electric double layers.
\end{abstract}



\maketitle


\section{Introduction}
The study of charged interfaces in electrolytes is a problem of fundamental importance to biophysics, electrochemistry, and colloidal science. \cite{Levin:RPP:02,BKNNSS:PP:05,DM:CR:90,Messina:JPCM:09,FPPR:RMP:10,BKSA:ACIS:09}  A proper understanding of the electric double layer (EDL) structure is essential to predict the stabilization of colloidal dispersions and the properties of biological systems. Under appropriate physical and chemical conditions, charged interfaces display complex and counter-intuitive phenomena such as the charge inversion and like charge attraction, which attracts a great theoretical and experimental interest. \cite{FPPR:RMP:10} These phenomena have been extensively observed in different systems including DNAs, self-assembled membranes and colloidal particles. \cite{GBPP:PT:00,ALWW:PNAS:03,LG:Nat:97,TMV:SSC:08}

In the generally accepted (by chemists) Gouy-Chapman-Stern theory, \cite{Lyklema:book:95,Conway:book:99} an EDL is composed of an internal Stern layer, where some counter-ions are tightly bound to the charged interface, and an outer diffuse layer, where counter-ions exert thermal motions.  The ion distributions in the diffuse layer are usually calculated using the Poisson-Boltzmann (PB) theory. \cite{Gouy:JP:10,Chapman:PM:13} Being of mean field nature, PB ignores the excluded-volume effects as well as electrostatic correlations of ions.  It is popularly believed that PB fails in the presence of multivalent ions or highly charged interfaces. \cite{GNS:RMP:02,LLPS:PRE:02}   Various methods, including but not limited to, modified PB theories, integral equation theories, and density functional theory have been developed to describe physics beyond the Poisson-Boltzmann framework. For example, it is notable that the state-of-the-art classical density functional theory \cite{YWG:JCP:04,WL:ARPC:07,KKDEG:JCP:10} has incorporated the hard-sphere repulsion and electrostatic correlation up to high precision, yielding results that are in quantitative agreement with Monte Carlo simulations.

While most of previous studies on the EDL model interfacial charges as uniformly distributed over the macroion surfaces; in reality, the surface charges are better modeled as discrete particles. When colloidal particles were placed in electrolyte solutions, the surface chemical groups release hydrogen cations to the solvent,  resulting in negatively charged interfacial ions, whose strengths depend on the environment conditions.  As an extremal example, phospholipids in aqueous solutions can carry a variable charge between $-4e$ and $+1e$ under different physiological conditions. \cite{FT:JPCC:07,TMB:BJ:08,WM:JPCB:10}
Recently, there has been many works demonstrating  \cite{FT:JPCC:07,TMB:BJ:08,CF:PRE:09,WM:JPCB:10,MGLH:EL:02,MRHQ:JPCB:09,CF:JCP:10,nelson1975effect,aguilella2010simulation,faraudo2010ionic} that interfacial charge discreteness has an important influence on the microion distribution near flat interfaces. For instance,  Faraudo et al. \cite{aguilella2010simulation,faraudo2010ionic} discovered that discrete interfacial charges can lead to an inversion of selectivity observed in a protein channel in the presence of multivalent cations. 
These motivate a systematical study on the effect of discrete interfacial charges.

Charged objects immersed in electrolytes usually have much lower dielectric constants than water.  Therefore polarization charges (image charges) on the interfaces are a relevant issue in the study of EDLs. The understanding of  image effects is of recent interest for electrostatic interactions of soft matter systems. \cite{JNPKP:PRL:08,KNFP:PRE:11,WM:JCP:09, WM:JPCB:10,WM:JCP:10,DBL:JCP:11} Counter-ions that are attracted to the interface are repelled by their likely charged image charges when they approach the interface.  Therefore image charges reduce the ion density in the vicinity of charged interface. This effect however diminishes as the surface charge strength increases,  as shown by Torrie {\it et al.} \cite{TVP:JCP:82} long time ago.  For a strongly charged surface, image charges push the whole EDL outward by a small distance, but otherwise has no significant influence on the phenomena such as the charge inversion, see recent works by Wang and Ma. \cite{WM:JCP:09, WM:JPCB:10,WM:JCP:10}  For recent reviews on image charge effects in colloidal and biological systems, see also references. \cite{HL:SM:08,XC:SIREV:11} We note, however, there is a recent work by Boda {\it et al.} \cite{BVENHG:JCP:06} showing that image charge effects may play important roles in biological systems.

If the interfacial charges are discrete, they also have image charges.  In aqueous solvents and for planar interfaces, these images are almost identical to the source charges and therefore essentially double the surface charge density.  This may lead to substantial modification of the EDL structure, as we shall show in the present work.  Finally, charged interfaces are usually not smooth at microscopic scales.  Surface roughness at atomic scales may interfere with image charge effects, and therefore change the physics of the charge inversion phenomena.  It is our purpose to study the interplay between image charges, interfacial charge discreteness, and surface roughness using a model system of a spherical colloid.

To study image charge effects for a generic interface, it is necessary to numerically solve the corresponding boundary value problem for the Poisson equation.  This is usually too time-consuming to be feasible in Monte Carlo simulations.  For some of recent works on smooth interfaces, see references. \cite{LZHM:CCP:08,BGNHE:PRE:04}   In this work, we shall explore the effects of image charges, interfacial charge discreteness, and surface roughness on the zeta potential and the charge inversion phenomenon for spherical geometries by using a recent method of multiple images. \cite{CDJ:JCP:07,GX:PRE:11}  We shall compare three different toy models for charged interfaces: a smooth surface with uniform surface charge density, a smooth surface with discrete interfacial charges, and a rough surface with discrete interfacial charges. This problem is difficult to address using the conventional method of the spherical harmonic expansion, \cite{Messina:JCP:02,RL:JCP:08} due to the intensive computation cost of the polarization potential.  This difficulty can be surmounted using the recently developed method of multiple images. \cite{CDJ:JCP:07,GX:PRE:11}

The term {\em zeta potential}  \cite{Hunter:book:81,delgado2007measurement} is intimately related to the Smoluchowski theories for electrophoresis: It is defined as the electrical potential in the interfacial double layer (DL) at the location of the supposed {\em slipping plane} versus a point in the bulk fluid away from the interface.   The existence of a {\em slipping plane} is one of the fundamental assumptions of electrophoresis theories, but we are not aware of any direct experimental evidence for its existence. Defined as such, the zeta potential can not be directly measured, but can only be inferred indirectly from electrokinetic data, through the application of Smoluchowski theories. Numerical calculations for the zeta potential have been made under different kind of geometries and various electrolytes.\cite{goel2008structure,modak2011effect} For 1:1 electrolyte with low salt concentrations, it was shown that the mean-field PB theory has a very good agreement with the simulation results. But in other cases, for example, with a $2:2$ salt in electrolytes, the PB fails to predict the result both qualitatively and quantitatively.\cite{degr¨¨ve1993monte}

For strongly charged surfaces, there is a layer of counter-ions strongly bound to the dielectric interface.  If the inversion of the electrophoretic mobility occurs,  this condensed layer must move with the colloid in the electrophoresis. Diehl and Levin \cite{DL:JCP:06} argued that in numerical computations of the zeta potential using Monte Carlo simulations, the slipping plane should be identified at about one counter-ion diameter away from the colloidal surface. In the presence of the charge inversion, an alternative, but probably even more natural choice of the slipping plane would be the peak of the integrated charge distribution function (ICDF).  This later choice would rigorously identify the charge inversion with the reversal of the colloidal mobility, which is generally assumed to be true.  It is interesting to note that in most numerical simulations, the peak of the ICDF is indeed about one micro-ion diameter away from the surface.  We note further that near the charge inversion threshold, the precise location of the slipping plane is only of minor importance, because the potential profile changes very slowly near the peak of the ICDF. We shall follow the choice of Diehl and Levin \cite{DL:JCP:06} in this work.

The remaining of this work is organized as follows. In Section II, we  present the three distinct models for interfacial charges and the simulation method used in this work. In Section III, we present the simulation results and discuss in details the effects of interfacial charge discreteness, image charges and surface roughness.  Finally, we draw the concluding remarks in Section IV.

\section{Models and methods}

We consider a charged colloidal particle with radius $a =  2nm$ and dielectric constant $ \varepsilon_\mathrm{i} = 2$, hereafter referred as the macroion, immersed in a $2:2$ symmetric electrolyte.  Such a macroion can be used to model micelles, dendrimers and other colloids. \cite{schmitz1993macroions,hiemenz1997principles,huang2000counterion,popa2009long,guerrero2010overcharging}
The aqueous solvent is modeled as a dielectric continua with a dielectric permittivity $\varepsilon_\mathrm{o} = 78.3$, while the ions are modeled as small hard spheres with diameter $\tau = 0.4 nm$ and with all charges in their centers.  The spherical Wigner-Seitz (WS) cell model \cite{Linse:APS:05} is employed for the boundary in our simulation.  The macroion is located at the center of the cell (with radius $R$) and has a bare charge $Q_M=Z_{M}e$, surrounded by the solvent and microions.  The electrolyte is treated at the level of the restricted primitive model.  The microions are confined in the spherical WS cell.  There are $N_+$ counter-cations with valence $Z_+ = 2$ and $N_-$ co-anions with valence $Z_- = - 2$ in the system.  The whole system is charge-neutral, hence $N_+ Z_+ + N_- Z_- + Z_M = 0$.

\begin{figure*}[htbp!]
\begin{center}
\includegraphics[width=0.9\textwidth]{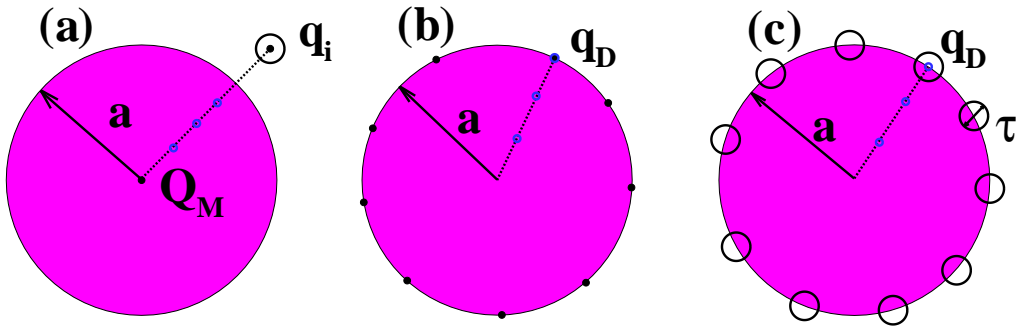}
\caption{Schematics for three models of interfacial charges studied in this work:
(a) SURF1: uniform surface charges.  It is equivalent to a single point charge at the center of macroion;
(b)SURF2:  discrete interfacial point charges on the surface;
(c) SURF3: discrete interfacial charges with an excluded volume. Also in (a), we show the image charges of a mobile microion $q_i$ as blue empty circles. In (b) and (c), a discrete interfacial charge $q_D$ produces image charges.  The Kelvin image coincides with the source charge itself, while other images lie inside the sphere.}
\label{surfacecharge}
\end{center}\end{figure*}

To explore the effects of image charges, interfacial charge discreteness, and surface roughness on the structure of EDL, we shall systematically compare three different models for interfacial charges, which are illustrated in Fig. \ref{surfacecharge}.     In model SURF1, the interfacial charges are uniformly distributed on the colloidal surface.   Equivalently, we can also put all surface charges onto the center of the spherical colloid.   Because of the spherical symmetry, the image charges of all interfacial charges cancel each other, so they have no influence on mobile microions in the bulk.   In model SURF2, there are $N_D$ point charges of valence $Z_D$ distributed on the smooth spherical colloidal surface. The total bare charge of the macroion is therefore $Q_M=q_{D}N_{D}$.
{Three different valences of discrete interfacial charges are studied: $Z_D\in\{-1,-2,-3\}$. In practice, $Z_D = -1$ corresponds to the carboxylate functional group\cite{andr¨¦2004role}; $Z_D = -2$ can be some kind of phospholipid such as 2,2-dimethoxy-2-phenylacetophenone (DMPA); \cite{faraudo2007phosphatidic} finally the so-called PIP2 lipids \cite{CF:JCP:10,lorenz2008hydrogen,mclaughlin2002pip2} can have a valence between $-2$ and $-4$ depending on the solvent environment.}  The locations of these $N_D$ charges are determined by running an MC minimization of the electrostatic energy near zero temperature with the constraint that the charges remain on the colloidal surface.  Typically, after $10^{5}N_D$ MC moves, the interfacial charges form a Wigner crystal on the surface.   Model SURF3 is the same as model SURF2, with the only difference that each interfacial charge acquires an excluded volumes (bumps) with diameter $\tau_D = 0.4 nm$.  For simplicity of computations, the portions of these bumps outside the sphere are assumed to have the same dielectric constant as the solvent. These two models have also been used in the study of ion binding to polyelectrolytes. \cite{MGCRS:TCA:09}

The bumps on the interface in SURF3 may have two competing impacts.  Firstly they increase the minimal distance between interfacial charges and mobile ions, by an amount of $\tau_D/2 = 0.2nm  $.   This tends to weaken the interaction energy between interfacial charges and the mobile ions, and therefore suppress binding between them.  Secondly the bumps also provide more interfacial area, so that more than one counter-ions can be adsorbed near a given interfacial charge. The competition between these two tendencies plays an important role in the local structure of the EDL.

The electric potential distribution, $\Phi(\mathbf{r})$, for a snapshot ion distribution satisfies the Poisson equation (Gauss unit),
\begin{equation}
-\nabla\cdot\varepsilon(\mathbf{r})\nabla \Phi(\mathbf{r})=4\pi\sum_j\delta(\mathbf{r}-\mathbf{r}_j),
\end{equation}
subject to standard electrostatic boundary conditions.  The dielectric function $\varepsilon(\mathbf{r})$ takes $\varepsilon_\mathrm{i}$ inside the sphere and $\varepsilon_\mathrm{o}$ outside,  $\delta(\cdot)$ is the Dirac delta function, and the index $j$ runs over both the mobile microions in the bulk as well as the interfacial charges on the macroion surface.  Numerically solving this Poisson equation in three dimensional space is time-costly.  Luckily for the spherical geometry, there is an efficient image charge algorithm. \cite{CDJ:JCP:07,GX:PRE:11}

Given the dielectric boundary, the electrical potential at $\mathbf{r}$ generated by one unit point charge at $\mathbf{r}'$ is given by the electrostatic Green's function $G(\mathbf{r},\mathbf{r}')$ that satisfies the following differential equation,
\begin{equation}
-\frac{1}{4\pi \varepsilon_\mathrm{o}}\nabla\cdot\varepsilon(\mathbf{r})\nabla G(\mathbf{r},\mathbf{r}')=\delta(\mathbf{r}-\mathbf{r}'),
\end{equation}
with homogeneous electrostatic boundary conditions. We use light italic font $r'=|\mathbf{r}'|$ to represent the radial distance in spherical coordinates.  For interfacial charges $r' = a$ while for mobile ions $r' >a$. The Green's function is a linear superposition of the point-charge Coulomb potential in free space, \begin{equation}
G_0(\mathbf{r},\mathbf{r}')=\frac{1}{|\mathbf{r}-\mathbf{r}'|},
\end{equation}
and the potential of all image charges due to the polarization of the macroion $G_\mathrm{im}(\mathbf{r},\mathbf{r}')$ ( with a unit source charge fixed at $\mathbf{r}'$):
\begin{equation}
G(\mathbf{r},\mathbf{r}')=G_0(\mathbf{r},\mathbf{r}')
+G_\mathrm{im}(\mathbf{r},\mathbf{r}').
\label{G-G0-Gim}
\end{equation}
For the spherical geometry, the image potential was discussed in details in reference  \onlinecite{CDJ:JCP:07}.   Here we invoke the result directly:
\begin{equation}
G_\mathrm{im}(\mathbf{r},\mathbf{r}')=\frac{-\gamma a/r'}{|\mathbf{r}-\mathbf{r}_K|}+\int_0^{r_K}\frac{\gamma \upsilon (x/r_K)^{\upsilon-1}/a}{|\mathbf{r}-\mathbf{x}|}dx,
\end{equation}
where $\upsilon=\varepsilon_\mathrm{o}/(\varepsilon_\mathrm{i}+\varepsilon_\mathrm{o})$, $\gamma=(\varepsilon_\mathrm{i}-\varepsilon_\mathrm{o})/(\varepsilon_\mathrm{i}+\varepsilon_\mathrm{o}) \approx -1$, and $\mathbf{x}=x\mathbf{r}'/r'$.  The first term is due to a (likely charged) point image with charge $-\gamma a/r'$ at $\mathbf{r}_K=\mathbf{r}'a^2/r'^2$ (Kelvin image), while the second term is due to an (oppositely charged) line image extending from $\mathbf{r}_K$ back to the center of the sphere.  These image charges are overall neutral.  By contrast, a planar interface only produces a point image (Kelvin image), and is not charge-neutral.  The importance of the line image is controlled by the ratio between the Debye length and the sphere radius, $\lambda/a$.  For our system, this parameter is not big, neither is it negligibly small, see Table \ref{parameter}.

Using the $I$-point Gauss-Legendre quadrature to approximate the line integral, $G_\mathrm{im}(\mathbf{r},\mathbf{r}')$ can be rewritten as the potential due to a total of $I+1$ image charges \cite{CDJ:JCP:07}
\begin{equation}
G_\mathrm{im}(\mathbf{r},\mathbf{r}')=\sum_{m=0}^I\frac{q_m}{|\mathbf{r}-\mathbf{x}_m|}, \label{image}
\end{equation}
where
$q_m=\frac{\omega_m}{2}\frac{\gamma a}{r'}$, and locations $x_m=r_\mathrm{K}\left(\frac{1-s_m}{2}\right)^{1/\upsilon}$,
$\{\omega_m, s_m, 1\leq m\leq I\}$ are the $I$-point Gauss weights and locations on the interval $[-1,1]$.  Note that $m =0$ corresponds to the Kelvin image charge, i.e., we have $\omega_0=-2$ and $s_0=-1$.  In this simulation we choose $I = 2$.
The image charges for mobile and interfacial charges in models are illustrated in Fig. 1 (a) and (b)(c), respectively.

The total Hamiltonian of the system can be expressed as a sum of three contributions: \cite{GX:PRE:11}
\begin{equation}
U=\sum_{i=1}^{N} U_{i}^{\mathrm{Mm}}
+ \sum_{i=1}^{N} \sum_{j=i}^{N}U_{ij}^{\mathrm{mm}}
+ U_\mathrm{HS}. \label{hamiltonian}
\end{equation}
The first part $U_{i}^{\mathrm{Mm}}$ is the interaction between macroion and microions, while the second part $U_{ij}^{\mathrm{mm}}$ is the interaction between microions.  The third part $U_\mathrm{HS}$ is the hard sphere repulsions between mobile ions, interfacial ions, the macroion, and the WS cell shell.  It takes the positive infinity when any volume exclusion constraint is violated and zero otherwise.

For SURF1, the interfacial charges are uniformly distributed on the sphere, hence their image charges, when added up, cancel each other, and have no interaction with the mobile ions. Therefore  $U_{i}^{\mathrm{Mm}}$ is the direct Coulomb interaction between the central charge at the origin and the mobile ion,
\begin{equation}
\beta U_{i}^{\mathrm{Mm}}=l_BZ_MZ_iG_0(0,\mathbf{r}_i),
\end{equation}
where $l_B=e^2/4\pi\epsilon_0\varepsilon_\mathrm{o} k_BT$ is the Bjerrum length, $\epsilon_0$ is the vacuum permittivity, $k_BT$ is the thermal energy, and $\beta=1/k_BT$. For SURF2 and SURF3, $U_{i}^{\mathrm{Mm}}$ is the interaction between interfacial ions and the mobile ions,
\begin{equation}
\beta U_{i}^{\mathrm{Mm}}=\sum_{n=1}^{N_D}l_BZ_DZ_iG(\mathbf{r}_i,\mathbf{r}_n),
\end{equation}
where $G$ is given by Eq.~(\ref{G-G0-Gim}), while $\mathbf{r}_n$ is the position of the $n$th interfacial ion.  It should be noted that due to the symmetry of the Green's function, $G(\mathbf{r}_i,\mathbf{r}_n)=G(\mathbf{r}_n,\mathbf{r}_i)$,  the interactions between the interfacial ions and the image charges of the mobile ions have been already included.   Similarly, the microion-microion interaction (the second term $U_{ij}^{\mathrm{mm}}$ in Eq. \eqref{hamiltonian}) can be written as,
\begin{equation}
\beta U_{ij}^{\mathrm{mm}}=l_BZ_iZ_j\left[(1-\delta_{ij})G(\mathbf{r}_i,\mathbf{r}_j)
+\frac{\delta_{ij}}{2}G_\mathrm{im}(\mathbf{r}_i,\mathbf{r}_j)\right],
\end{equation}
where $\delta_{ij}$ is the Kronecker delta. When $i=j$, it represents the interaction of the charge and its images.

\begin{figure}[htbp!]
\begin{center}
\includegraphics[width=3.in]{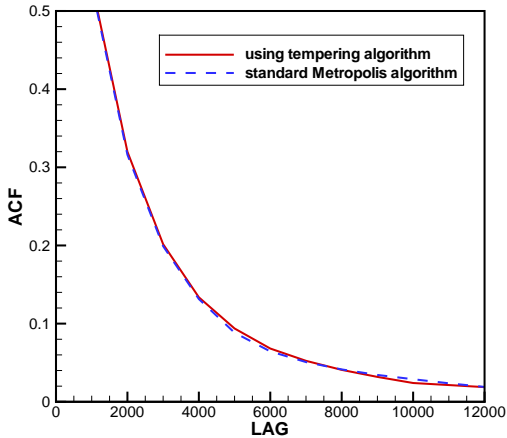}
\caption{Energy autocorrelation function (ACF) as a function of MC steps.
System parameters: the surface charge density:
$0.1C/m^{2}$ , $Z_{D}=-3$, salt concentration $C=0.0755M$. For other parameters, see Table \ref{parameter}.
}
\label{acf.fig}
\end{center}\end{figure}

Canonical-ensemble Monte Carlo simulations based on the standard Metropolis acceptantce/rejection rule \cite{Metropolis:53,FS:book:02} are carried out to obtain the equilibrium properties of the model systems. The initial configuration of each system is generated by randomly placing the ions into the simulation cell satisfying the constraints of the hard-core repulsion.  The total number of mobile ions, $N$, varies from about 200 to 300, depending on the salt concentrations and the WS cell radius.  In each simulation, we perform $1.2\times10^6N$ MC moves per particle. The first  $10^5N$ MC moves are performed using the tempering technique, \cite{marinari1992simulated,geyer1995annealing} where we start with $T=2100K$ (Bjerrum length $l_B=0.1nm$), and then slowly cool down to room temperature $T=298K$ (Bjerrum length $l_B=0.71nm$). This is followed by another $10^5N$ MC moves for equilibration.  Finally, $10^6N$ MC moves are performed to store the data for statistical analysis. In order to achieve high sampling efficiency, the acceptance ratio was kept about 0.3 by adjusting the maximum step size of ion motion. \cite{roberts1997weak}   We find that the autocorrelation function (ACF) of total energy typically decays to zero after about 12000 MC steps (for the whole system).  Since in our simulation, about
$200\times10^{6}$ MC steps are calculated, we have nearly 20000 independent samples.  The ACF for one typical run is illustrated in Fig.~\ref{acf.fig}.

We chose symmetric and multivalent $2:2$ electrolytes instead of more common asymmetric $2:1$ electrolytes mainly from the consideration of computational capacity.  Nevertheless, $2:2$ divalent salt systems are also experimentally relevant \cite{bastos1993effect,bastos1994colloidal} and have been studied both analytically \cite{lozada1999nonlinear,ennis1995dressed,ennis1996effective} and numerically. \cite{degr¨¨ve1998monte,deserno2001overcharging}   It is known that the structure of EDLs is mainly determined by the valence of counterions. \cite{goel2008structure,modak2011effect}  In our simulations, the salt concentrations of the systems take three values: $C=0.0755,$ $0.155$ and $0.485M$; the surface charge density of the macroion, $\sigma$, varies from $-0.1$ to $-0.6$ $Cm^{-2}$, correspondingly the bare charges range from $-30e$ to $-180e$, and the Gouy-Chapman length
($l_{GC}=e/2\pi Z_+ l_B\sigma$) ranges from 0.031 to $0.19nm$.  The radius of the WS cell $R$ is chosen to be at least 14 times longer than the Debye length $\lambda$, so that the influence of the cell wall on the EDL structure is negligible.  The radius of the spherical macroion is kept $a=2nm$. All simulation parameters are summarized in Table \ref{parameter}.

\begin{table}[h]\begin{center}
\renewcommand{\arraystretch}{1.3}
\caption{ Relevant system parameters used in the Monte Carlo simulations}  \label{parameter}
\vspace{5mm}
\begin{tabular}{ll}\hline\hline
$\varepsilon_\mathrm{i}=2$   &    Colloidal dielectric constant \\
$\varepsilon_\mathrm{o}=78.3$   &    Solvent dielectric constant \\
$Z_M=-30\sim-180$   &   Macroion valence \\
$Z_\pm=\pm 2$   &    Counterion and coion valence \\
$Z_D=-1, -2, -3$        & Interfacial ion valence      \\
$a=2nm$        &   Macroion radius    \\
$\tau=0.4nm$   &   Microion diameter     \\
$\tau_D=0.4nm$ &   Diameter of interfacial ions in SURF3    \\
$T=298 K$      & Room temperature      \\
$l_B=0.71nm$   & Bjerrum length      \\
$l_{GC}=0.19\sim 0.031nm$ & Gouy-Chapman length\\
$C=0.0755, 0.155, 0.485 M$   & Three salt concentrations      \\
$\lambda=0.54, 0.38, 0.22 nm$   & Corresponding Debye lengths  \\
$R=7.72, 6.13, 4.19nm$ & Corresponding WS cell radii \\    \hline\hline
\end{tabular}
\end{center}
\end{table}

The zeta potential and the integrated charge distribution function (ICDF) are calculated for three different surface charge models, SURF1, SURF2 and SURF3.   As discussed in the Introduction, the zeta potential is defined as the average potential one micro-ion diameter away from the colloidal surface, $\zeta=\Phi(a+\tau)$.   For spherical EDLs, this potential can be explicitly obtained by integrating the Poisson equation: \cite{YWG:JCP:04}
\begin{equation}
\zeta=\frac{4\pi}{\varepsilon_\mathrm{o}}\int_{a+\tau}^{\infty}\sum_i\rho_i(r)Z_ie\left(r-\frac{r^2}{a+\tau}\right)dr, \label{zeta}
\end{equation}
where $\rho_i(r)$ is the mean density of the $i$th ion species, and $Z_i$ its valence. The integrated charge distribution function (ICDF) $Q(r)$ as a function of radius $r$ is given by
\begin{equation}
Q(r)=Q_M+[Z_+N_+(a,r)+Z_-N_-(a,r)],
\end{equation}
where $N_\pm(a,r)$ are the average numbers of positive/negative ions in the spherical shell between $a$ and $r$, and $Q_M$ the bare charge of macroion. It is found that $Q(r)$ changes sign at $r \approx a + 3 \tau/4$  when the charge inversion takes place.  The maximum of the ICDF curve is defined as the inverted charge, which equals zero if no charge inversion happens. See Fig.~\ref{surf1_icdf}, Fig.~\ref{surf2_ICDF} and Fig.~\ref{surf3_icdf} for illustrations of ICDF curves.  We have also calculated the standard deviation of the zeta potential to make sure that the simulation results are accurate.  For the system parameters shown in Fig.~\ref{acf.fig}, for example, the zeta potential is found to be  $\zeta = 31.833mV$, while its standard deviation is: $0.279mV$.

\section{Results and discussion}

\subsection{SURF1: Uniform surface charge distribution}

\begin{figure*}[htbp!]\begin{center}
\includegraphics[width=7in]{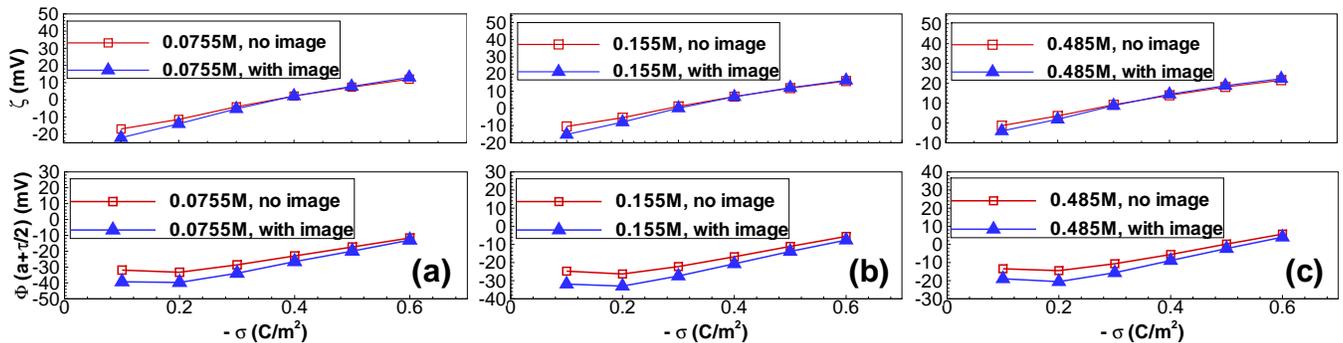}
\caption{Top row: The zeta potential (average potential at one micro-ion diameter away from the interface) for SURF1 as the function of surface charge density $\sigma$ for both with and without image charge effects. Three different concentrations of 2:2 salt are calculated: (a) $C=0.0755M$, (b) $C=0.155M$, and (c) $C=0.485M$.  Bottom row: average potential at half micro-ion diameter away from the interface.  }
\label{surf1}
\end{center} \end{figure*}

In model SURF1, interfacial charges are smoothly distributed on the spherical colloidal surface.  The image charge potential of interfacial charges is averaged out and has no influence, but microions do have image charges that are inside the colloidal sphere.  These images repel the source charge away from the colloidal surface, resulting in a depletion zone for mobile microions near the interface.  It has been shown that except this depletion effect, image charges of the mobile ions have no influence on the charge inversion phenomena. \cite{Messina:JCP:02} In agreement with these results, our simulation results show that the image charges only slightly change the value of the zeta potential.  Illustrated in the first row of Fig. \ref{surf1} are the zeta potential for model SURF1 with and without image charges, for three different concentrations of 2:2 salt.

\begin{figure}[htbp!]\begin{center}
\includegraphics[width=0.53\textwidth]{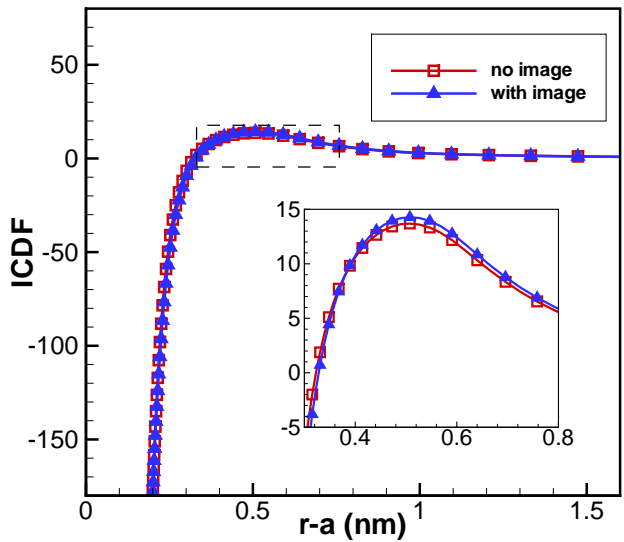}
\caption{Model SURF1. The integrate charge distribution functions (ICDFs). $\sigma=-0.6C/m^2$ and $C=0.155M$.  The peak of these curves are at slightly more than one micro-ion diameter outside the surface.  }
\label{surf1_icdf}
\end{center} \end{figure}

For comparison we also show in the bottom row of  Fig.~\ref{surf1} the average potential $\Phi(a + \tau/2)$ at distance of one microion {\it radius} $\tau/2$ away from the interface (which is often used in literature as the zeta potential).  We find that this potential is drastically different from $\Phi(a + \tau)$. More importantly, $\Phi(a + \tau/2)$ remains negative for the whole range of the surface charge strengths studied for the cases of salt concentration $0.0755M$ and $0.155M$, and therefore shows no sign of the charge inversion.  By strong contrast, the  potential $\Phi(a + \tau)$ indicates the charge inversion at $\sigma \approx - 0.35Cm^{-2}$ (salt $0.0755M$) and  $\sigma \approx - 0.3Cm^{-2}$ (salt $0.155M$).   The plot for the ICDFs in Fig.~\ref{surf1_icdf} also clearly shows the charge inversion for $C = 0.155$ and $\sigma = -0.6C/m^2$.  We therefore conclude that $\Phi(a + \tau)$, rather than $\Phi(a + \tau/2)$, is a good definition of the zeta potential in term of the characterization of the charge inversion phenomenon.  This agrees with the results by Diehl and Levin. \cite{DL:JCP:06}



\subsection{SURF2: Discrete surface charges without volume effect}
\begin{figure*}[htbp!]\begin{center}
\includegraphics[width=7in]{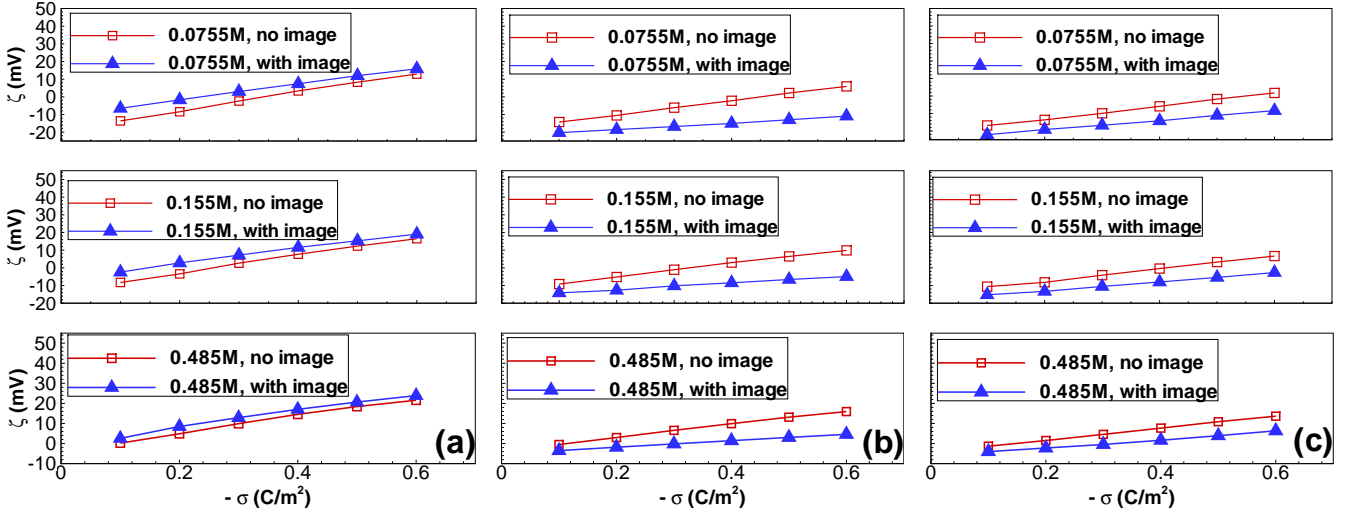}
\caption{The zeta potential for SURF2 as a function
of mean surface charge density, with various valences of interfacial charges, both with and without image effects.  Left: $Z_D=-1$; middle: $Z_D=-2$; right: $Z_D=-3$.  It is clear from these plots that image charges enhance charge inversion for $Z_D = -1$, and suppress charge inversion for $Z_D = -2, -3$. 
}
\label{surf2}
\end{center}\end{figure*}

In model SURF2,  interfacial charges are point charges distributed on the sphere.   The corresponding zeta potential as a function of surface charge density is illustrated in Fig. \ref{surf2}.  Three cases for the valences of interfacial charges are shown, with $Z_D=-1, -2$ and $-3$, respectively.
For $Z_D=-1$ (Fig.~\ref{surf2},  left column),  the zeta potentials with image charge effect become substantially higher than those with no image charge effect.  Comparing with Fig.~\ref{surf1},  it is evident that the image charge effects greatly enhance the overcharging tendency.
This is clearly due to the effective doubling of interfacial charge valence by their image charges.

\begin{figure}[htbp!]\begin{center}
\includegraphics[width=0.53\textwidth]{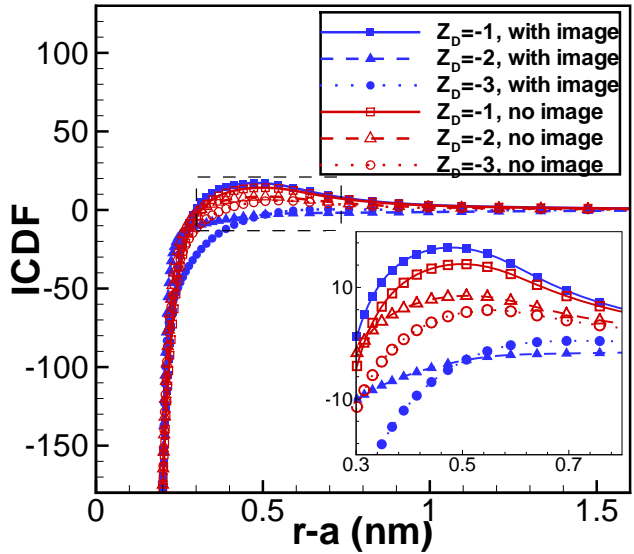}
\caption{Model SURF2. The integrate charge distribution functions (ICDFs) for different valences of interfacial ions. $\sigma=-0.6C/m^2$ and $C=0.155M$.} \label{surf2_ICDF}
\end{center} \end{figure}

Interestingly, in the cases of higher valences of interfacial ions, $Z_D=-2$ (Fig.~\ref{surf2},  center column) and $Z_D = -3$ (Fig.~\ref{surf2},  right column), image charge effects influence zeta potential profiles in the opposite direction.  The zeta potentials with the image charges remain uniformly {\it below} those without images, therefore the tendency of the charge inversion is substantially suppressed by the image charge effects.  Furthermore, the image charge effects grow with increasing surface charge density, which is the opposite of what was discovered for uniformly charged surfaces by Torrie {\it et al.}. \cite{TVP:JCP:82} It is also interesting to note that the suppression of the charge inversion is more significant for $Z_D = -2 $ than for $Z_D = -3$.

These effects of  image charges are also evident in the ICDF plots.  In Fig. \ref{surf2_ICDF} we show ICDF curves for various interfacial charge valences, with fixed surface charge density $\sigma=-0.6C/m^2$ and salt concentration $C=0.155M$.  It is clear that for $Z_D = -1$, the integrated charges with image charges always lie above that without image charges, while for $Z_D = -2$ and $-3$, the curves with image charges lie below those without image charges.  In fact, for these latter two cases, the image charge effects completely eliminate charge inversion.
The suppression is again more pronounced for $Z_D = - 2$ than for $Z_D = - 3$.

\begin{figure*}[htbp!]
\begin{center}
\includegraphics[width=2.3in]{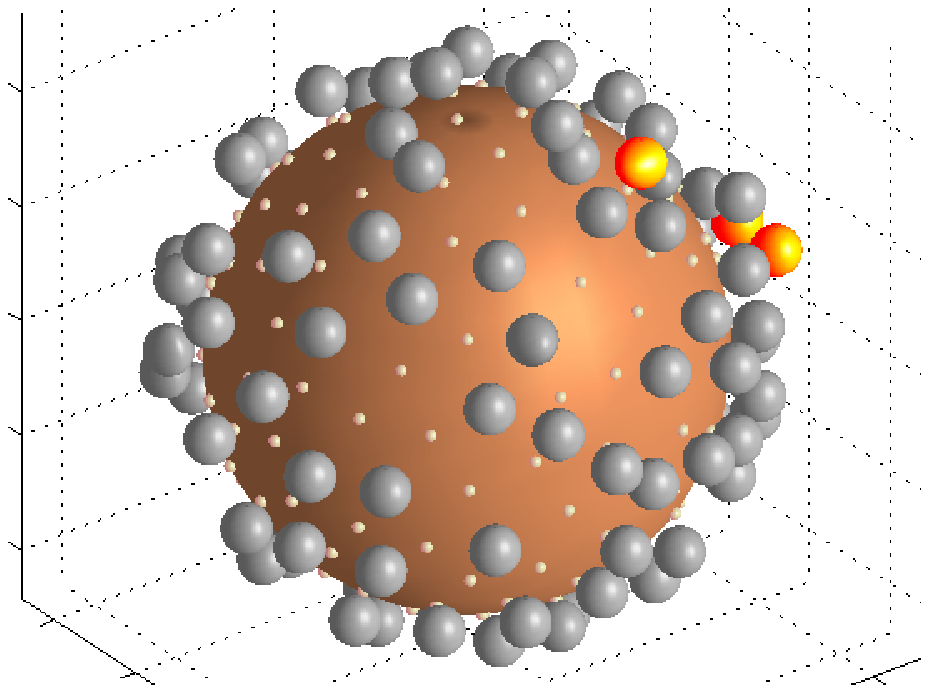}%
\includegraphics[width=2.4in]{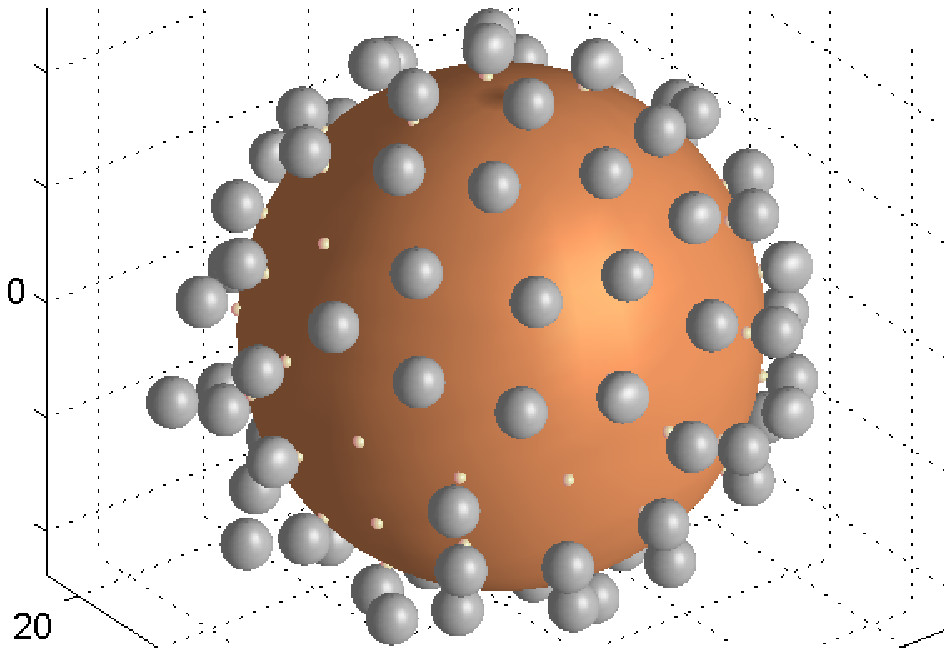}%
\includegraphics[width=2.2in]{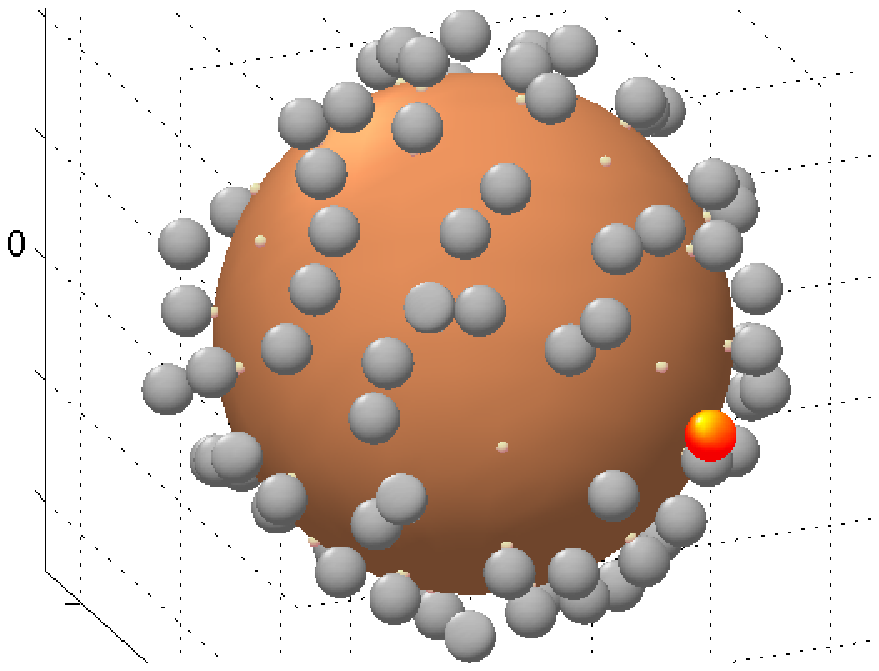}%
\end{center}
\caption{
Typical snapshots of model SURF2 after equilibrium is achieved.
Gray spheres: counterion; yellow spheres: coion; small white
spheres: interfacial charges.
Left: $Z_D= - 1$; Center:  $Z_D = -2$;  Right: $Z_D= - 3$.  System parameters:  salt concentration $0.155M$, surface charge density $0.6 C/m^2$. }
\label{snapshot3}
\end{figure*}

The huge suppression of the charge inversion for divalent interfacial charges is likely due to the strong binding between these interfacial charges and the divalent counter-ions, which completely neutralizes the interfacial charge groups.  Other counter-ions can no longer be adsorbed onto the nearby area if such a binding occurs.  The binding energy is about $2Z_D \ell_B/(\tau/2) \approx 14 k_B T$ without image charges and is further doubled by the image charge effects.  This binding may strongly decreases the effective surface charge density and therefore suppresses charge inversion.  To verify this physical picture, we have looked at typical simulation snapshots of ion distributions near the colloidal surface, for model SURF2 (no bumps, with image charges).  One typical simulation snapshot is shown in Fig.~\ref{snapshot3} for each of three cases $Z_D = -1,-2,-3$.  It is found that for $Z_D = -2$ (Fig.~\ref{snapshot3}, center panel), about $90\%$ of interfacial charges are closely bound to a counter-ions, and become completely neutralized.  Such a binding is clearly strengthened by image charge effects.

By contrast, for $Z_D = -1$ (Fig.~\ref{snapshot3}, left panel), about $46\%$ of interfacial charges are closely bound to counter-ions.  Since the counter-ions carry charge $2e$ while interfacial charges carry $-e$, each of these bindings contribute to the charge inversion.  Indeed model SURF2 with $Z_D = -1$ goes charge reversal at a much lower surface charge density, comparing with SURF1, as one can see from Fig.~\ref{surf2} and Fig.~\ref{surf1}.  Finally, for $Z_{D}= - 3$ (Fig.~\ref{snapshot3}, right panel), we can see that the majority of interfacial charges are also bound to counter-ions.  In this case, however, one-to-one binding does not invert the charge, nor does it completely neutralize the interfacial charge group.  Some interfacial charges actually can bind to two counterions. Their numbers are however not enough to invert the charge of the whole colloid.   Model SURF2 with $Z_D = -3$ goes charge reversal at higher surface charge density than model SURF1.

We note that the binding between interfacial charges and counterions are important for $Z_D = -2, -3$ even in the absence of image charges.  Image charge effects however substantially enhance these bindings.  Our simulation demonstrates the following: comparing with a uniform surface charge distribution, the interfacial charge discreteness enhances charge inversion if interfacial charge groups have smaller valence than counterions, and suppresses charge inversion if interfacial charge groups have equal or larger valence than counterions.  Image charges strengthen these effects by effective doubling of the surface charge density.



\subsection{SURF3: Discrete surface charges with finite exclusion volumes}

\begin{figure*}[htbp!]\begin{center}
\includegraphics[width=7in]{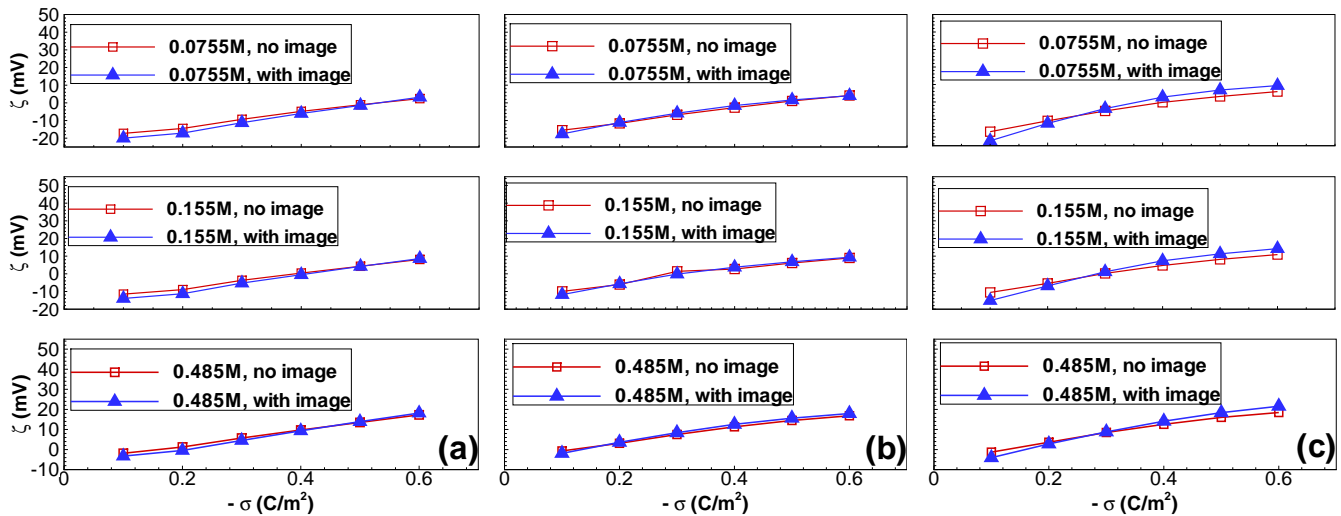}
\caption{The zeta potential for SURF3 as a function of surface charge density for different valences of interfacial charges in cases both with and without image charges. The valence of interfacial ions $Z_D=-1, -2, -3$ from left to right.} \label{surf3}
\end{center}\end{figure*}

\begin{figure}[htbp!]
\begin{center}
\includegraphics[width=0.53\textwidth]{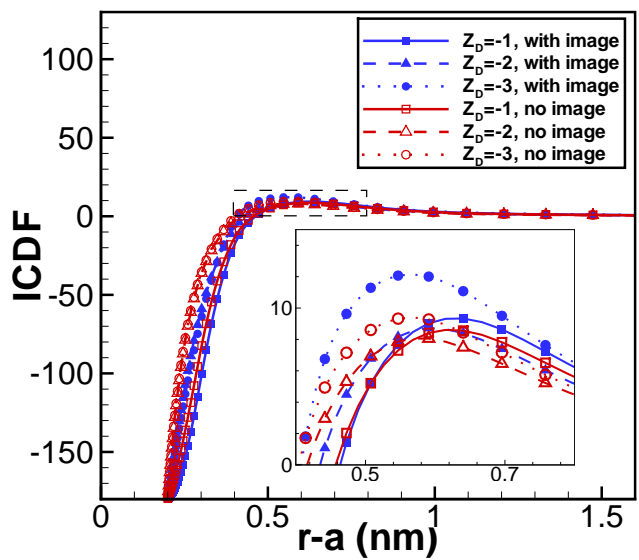}
\caption{Model SURF3.  The ICDF curves for $Z_D=-1,-2$ and $-3$ for the cases with and without image
charges. Salt concentration $C=0.155M$ and surface charge density $\sigma=-0.6C/m^{2}.$ } \label{surf3_icdf}
\end{center}\end{figure}

In model SURF3, there is a spherical excluded volume (bump) with diameter $\tau_D = 0.4nm$ around each interfacial charge, which prevents the counter-ions from getting closer than $\tau = 0.4nm$ to the interfacial charge.  Zeta potentials for SURF3 for various values of $C$ and $Z_D$ are shown in Fig.~\ref{surf3}. Comparing with Fig.~\ref{surf2}, we see that the effects of image charges become much smaller, in fact almost negligible, in the presence of volume exclusion effects of bumps.  Nevertheless, for all three cases of interfacial charge valences, image charge effects slightly {\em enhance} charge inversion, in strong contrast with model SURF2.  Furthermore, while in SURF2, the charge inversion threshold of the surface charge density increases with the valence of interfacial charges, in SURF3, the charge inversion threshold actually {\em decreases} with the valence of interfacial charges.
Another striking effect is that, the more realistic model SURF3 gives a much higher zeta potential than model SURF1, as one can see by comparing Fig.~\ref{surf3} with Fig.~\ref{surf1}.  These surprising results are probably due to the extra interfacial area around the bumps, where multiple counter-ions can be attracted by a given surface charge.  This effect becomes more important as the valence of interfacial charge increases.

We also plot the ICDF curves in Fig. \ref{surf3_icdf} for fixed salt concentration $C=0.155M$ and the surface charge density $\sigma=-0.6C/m^{2}$.   The figure shows the charge inversion is moderately enhanced by the image charges by examining the strength of the inverted charges.  The peaks of ICDF curves are also pushed a little further away from the surface by the image charges, in agreement with previous studies by other groups. \cite{TVP:JCP:82,HL:SM:08,WM:JCP:09, WM:JPCB:10,WM:JCP:10}



\begin{figure*}[htbp!]
\begin{center}
\includegraphics[width=7in]{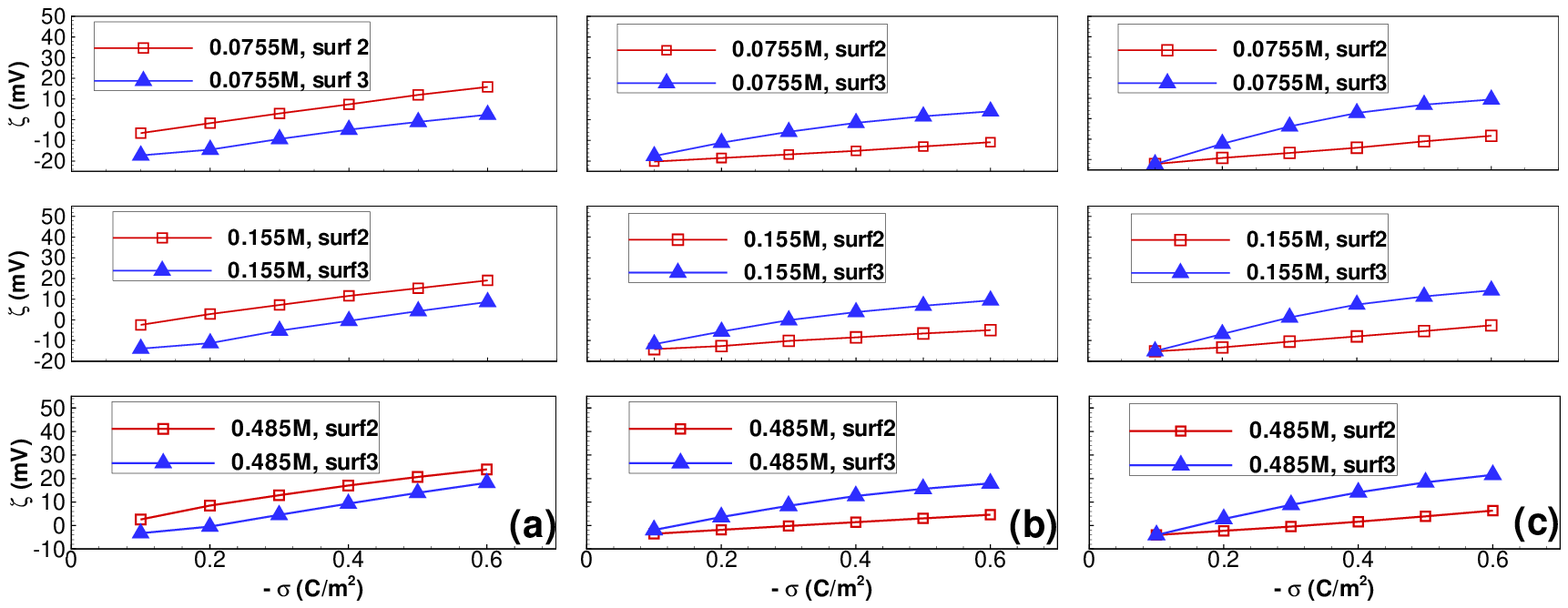}
\caption{The zeta potential for SURF2 and SURF3 (both with image charge effects) as a function of mean surface charge density. The valence of interfacial ions $Z_D=-1,-2,-3$ from left to right.}
\label{surf23}
\end{center}
\end{figure*}

We also  compare the zeta potential of models SURF2 and SURF3 (both with image charges) in Fig.~\ref{surf23}.  We find that for $Z_D = -1$, the data for SURF3 are uniformly and substantially lower than those for SURF2, indicating that charge inversion is strongly suppressed  by the excluded volumes.  By strong contrast, for  $Z_D = - 2$ or $- 3$, the zeta potential for SURF3 are substantially higher than those for SURF2, indicating that charge inversion is strongly enhanced by the bumps.

What is the physics underlying these huge differences between SURF2 and SURF3?  Recall that in model SURF3, each surface charge group has an exclusion volume (``bump") with radius $2 \AA$.  These bumps have two competing effects: Firstly the minimal distance between surface ions and the ions in the bulk is $4\AA$ in SURF3, instead of $2 \AA$ in SURF2.  Hence the maximal interaction energy between surface ions and counterions are approximately reduced by half in SURF3.   This seems to be the dominant factor in the case of $Z_D = -1$, leading to strong suppression of charge inversion in SURF3.   Secondly the bumps provide extra adsorption area for the counterions.   If the interaction between interfacial charges and counterions are already strong enough, these extra areas will lead to more counterions condensed near the interface, and therefore enhance charge inversion.  This seems to be the dominant factor for the cases of $Z_D = -2$ and $Z_D = -3$.

\section{Concluding remarks}
We have studied the effects of image charges, interfacial charge discreteness, and surface roughness on the zeta potential profile and the integrated charge distribution function for a strongly charged spherical colloid.  Such a study is possible owing to the recent development of image charge methods for a spherical boundary, which approximates the image potential with a few image point charges. Our main results are summarized as below:

\begin{enumerate}
\item[(1)]
In agreement with Diehl and Levin, \cite{DL:JCP:06} we find that the zeta potential defined at about one counter-ion diameter away from the colloidal interface provides a good indicator for the charge inversion phenomenon.


\item[(2)] The effects of image charges depend crucially on the nature of the surface charge distribution.  For uniform surface charges (SURF1), the influence of image charges on zeta potential is minor, in agreement with previous results by other groups.  For discrete surface charge groups with no excluded volumes (SURF2), we find that image charges strongly enhance charge inversion if the surface charge groups are monovalent, and strongly suppress charge inversion if the surface charge groups are multi-valent.  For discrete surface charge groups with excluded volumes (SURF3), we find that the effects of image charges only slightly enhance charge inversion.



\item[(3)] Model SURF3 with discrete interfacial charges and finite excluded volumes gives a much higher zeta potential than model SURF1, where the interfacial charges are continuous and the colloidal surface is smooth.  Finally, the effects of excluded volumes (bumps) of interfacial charges are substantial and depend crucially on the valences of surface charges.  For monovalent surface charges, we find that a high surface roughness strongly suppresses charge inversion, while for multi-valence surface charges, a high roughness strongly enhances charge inversion.   These results show that the roughness of charged interfaces is an extremely important factor.

\end{enumerate}

Overall, our simulation results demonstrate intricate and competing effects associated with image charges, interfacial charge discreteness, and surface roughness.  All these factors can substantially affect the profile of the zeta potential and the charge inversion.  Therefore short scale details of charged interfaces need to be better clarified before the structure of EDLs can be understood properly.



\section*{Acknowledgements}

The authors acknowledge the financial support from the Natural Science Foundation of China (Grant Numbers: 11101276, 11174196, and 91130012) and Chinese Ministry of Education (NCET-09-0556). The author gratefully acknowledges the (anonymous) referees' valuable comments which lead to an improvement of this paper.


\end{document}